# Few-cycle, Broadband, Mid-infrared Optical Parametric Oscillator Pumped by a 20-fs Ti:sapphire Laser

S. Chaitanya Kumar[1], A. Esteban-Martin[1], T. Ideguchi[2], M. Yan[2,3], S. Holzner[2], T.W. Hänsch[2,3], N. Picqué[2,3,4] and M. Ebrahim-Zadeh[1,5,*]

1. ICFO-Institut de CienciesFotoniques, Mediterranean Technology Park, 08860 Castelldefels, Barcelona, Spain;
2. Max-Planck InstitutfürQuantenoptik, Hans-Kopfermann-Strasse 1, D-85748 Garching, Germany;
3. Ludwig-Maximilians-Universität München, Fakultät fur Physik, Schellingstr. 4/III, 80799 Munchen, Germany;
4. Institut des Sciences Moléculaires d'Orsay, CNRS, Bâtiment 350, Université Paris-Sud, 91405 Orsay, France;
5. Institucio Catalana de RecercaiEstudisAvancats (ICREA), PasseigLluisCompanys 23, Barcelona 08010, Spain;
\* Corresponding author(s): e-mail: majid.ebrahim@icfo.es

**Abstract:** We report a few-cycle, broadband, singly-resonant optical parametric oscillator (OPO) for the mid-infrared based on MgO-doped periodically-poled LiNbO$_3$ (MgO:PPLN), synchronously pumped by a 20-fs Ti:sapphire laser. By using crystal interaction lengths as short as 250 µm, and careful dispersion management of input pump pulses and the OPO resonator, near-transform-limited, few-cycle idler pulses tunable across the mid-infrared have been generated, with as few as 3.7 optical cycles at 2682 nm. The OPO can be continuously tuned over 2179-3732 nm by cavity delay tuning, providing up to 33 mW of output power at 3723 nm. The idler spectra exhibit stable broadband profiles with bandwidths spaning over 422 nm (FWHM) recorded at 3732 nm. We investigate the effect of crystal length on spectral bandwidth and pulse duration at a fixed wavelength, confirming near-transform-limited idler pulses for all grating interaction lengths. By locking the repetition frequency of the pump laser to a radio-frequency reference, and without active stabilization of the OPO cavity length, an idler power stability better than 1.6% rms over >2.75 hours is obtained when operating at maximum output power, in excellent spatial beam quality with TEM$_{00}$ mode profile.

## 1. Introduction

The generation of coherent ultrashort light pulses with broad spectral bandwidth is of great interest for applications in optical metrology and frequency synthesis [1].In the mid-infrared (mid-IR) spectral range, such broadband ultrafast sources are of particular interest for high-precision frequency comb spectroscopy [2]. With the limited availability of conventional mode-locked lasers in the mid-IR, nonlinear frequency down-conversion techniques represent an attractive approach to the generation of ultrashort pulses in this region, inherently preserving the phase coherence properties of the input laser pump sources. A particularly effective approach is to exploit ultrafast optical parametric oscillators (OPOs) synchronously pumped by mode-locked femtosecond laser oscillators[3].By deploying Kerr-lens-mode-locked (KLM) Ti:sapphire and $Cr^{2+}$:ZnSe lasers, or femtosecond Yb-, Er- and Tm-fiber lasers, the potential of ultrafast OPOs for the generation of near-to mid-IR radiation has already been extensively demonstrated [4-10]. By exploiting degenerate doubly-resonant oscillator (DRO) configuration [4, 6-9], or singly-resonant oscillator (SRO) design [5,10], broadband radiation in the near- to mid-IR across spectral regions from ~1.5 to ~6 μm has been generated using quasi-phase-matched (QPM) nonlinear materials of MgO-doped periodically-poled $LiNbO_3$ (MgO:PPLN) and orientation-patterned GaAs (OP-GaAs), or birefringent crystals, $BiB_3O_5$ and$CdSiP_2$. The potential of such OPOs has also been demonstrated for applications in frequency metrology [5,7] and spectroscopy [6,8].

For the attainment of broadband output spectrum, the degenerate sub-harmonic DRO[4] offers an attractive approach, because of the convenient availability of large gain bandwidths in the vicinity of degeneracy under type 0 ($e \rightarrow ee$) interaction in QPM materials, or type I ($e \rightarrow oo$) phase-matching in birefringent crystals [9]. In addition, the sub-harmonic DRO can provide an output spectrum at exact degeneracy, where the signal and idler are mutually phase-locked, and also inherently phase-locked to the pump [4,7].As such the degenerate output can be used as a first step in frequency comb generation, subject to further active stabilization. However, an intrinsic feature of the sub-harmonic DRO is the spectral confinement of output to wavelength regions close to degeneracy, hence requiring pump sources at different wavelengths to access different spectral regions in the mid-IR. On the other hand, for arbitrary spectral coverage away from degeneracy with smooth and continuous tuning, the SRO configuration can be a desirable approach. In addition, the SRO scheme enables the deployment of widely established and more readily available femtosecond pump sources, such as the KLM Ti:sapphire laser, to access wide spectral regions in the mid-IR. The SRO configuration is also characterized by inherent passive stability, resulting in a smooth, uniform, and stable output spectrum with a well-defined structure and content across the full synchronous range of cavity delay, and provides excellent temporal and output power stability across the entire wavelength coverage of the OPO.

At the same time, in both sub-harmonic DRO and SRO configurations, for the attainment of largest output spectrum it would be desirable to deploy the broadest input pump bandwidth, and hence the shortest transform-limited pump pulse duration. In this respect, the KLM Ti:sapphire laser also represents the most viable and well-established source of ultrashort femtosecond pulses, capable of delivering the shortest transform-limited optical pulses with broadest spectral bandwidth[11]. When used as the pump in combination with the SRO design, it is potentially capable of generating few-cycle optical pulses with broad, stable and well-controlled bandwidth across the entire mid-IR spectral range of ~1 to 4.5 μm using MgO:PPLN as the nonlinear material. On the other hand, the deployment of ultrashort pump pulses presents additional challenges, both in terms of dispersion control of the input pump pulses as well as careful management of dispersion within the SRO cavity [12], to achieve the shortest output pulses, and the most stable broadband mid-IR idler output. Recently, a sub-harmonic DRO based on MgO:PPLN and synchronously pumped by 85-fs pulses from an Er-fiber laser at 1560 nm was reported, providing broadband 5-cycle (50 fs) optical pulses at exact degeneracy centered at 3120 nm[13].Here we report a femtosecond OPO in SRO configuration capable of providing few-cycle optical pulses tunable across mid-IR, from a center wavelength of 2179 nm to 3732 nm. The OPO is pumped by a 20-fs KLM Ti:sapphire laser at 790 nm, and provides clean, uniform, and well-defined broadband idler spectra at all operating wavelengths. By careful dispersion control of input pump pulses, OPO crystal and mirror coatings, as well as intra-cavity group delay compensation, idler pulses of 3.7 optical cycles have been generated at 2282 nm. The OPO can deliver broad idler spectra in the mid-IR, with a full-width at half-maximum (FWHM) bandwidth spanning over 422 nm at 3732 nm. With the Ti:sapphire repetition frequency locked to a radio frequency (RF) reference by servo-control of the cavity length, and without active OPO cavity length stabilization, we obtain a passive idler output power stability better than 1.6% rms over >2.75 hours, with excellent spatial beam quality in a $TEM_{00}$ mode profile.

## 2. Experimental setup

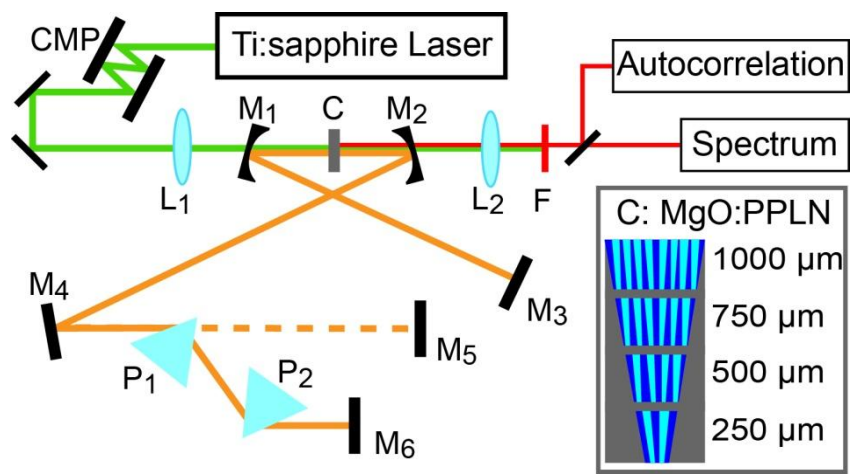

Fig. 1.Schematic of the few-cycle optical parametric oscillator. CMP: chirp mirror pair, M: mirrors, L:Lens, P: Prisms, F: Filter.

The schematic of the experimental setup is shown in Fig. 1. The pump source is a KLM Ti:sapphire oscillator providing up to 1 W of average power in ~20 fs pulses at the output of the laser at a repetition frequency (*frep*) of 100 MHz. In order to preserve the minimum pulse duration at the input to the nonlinear crystal inside the OPO, we use a pair of chirped mirrors (CMP) to pre-compensate the pump pulses for the dispersion caused by the focusing lens, $L_1$, and the input mirror, $M_1$. After four bounces through the CMP, corresponding to-80 $fs^2$/bounce, the measured pump pulse duration and the spectrum at the input to the nonlinear crystal are recorded, with the results shown in Fig. 2. The interferometric autocorrelation measurement resulted in pulse duration of 26 fs ($sech^2$), corresponding to ~10 optical cycles, with a spectral bandwidth of 39 nm (FWHM) centered at 790 nm. Using a $f$=75 mm focal length lens, the pump beam is then focused to a waist radius of $w_0$~50 μm inside the nonlinear crystal, which is a 1-mm-thick MgO:PPLN chip, with four different interaction lengths, $l$=250, 500, 750, 1000 μm, as shown in Fig. 1.Each interaction length is composed of a fanned grating structure, with QPM grating periods ranging from $\Lambda$ =19 to $\Lambda$=21.3 μm. The grating periods are optimally designed to maximize the gain bandwidth over 1600-3500 nm in the mid-IR, with the calculated idler bandwidth ranging from >1500 nm for $l$=250 μm to >270 nm for $l$=1000 μm, for a fixed grating period, $\Lambda$=20 μm. This custom design of the nonlinear crystal enables us to study the effect of interaction length on the pulse duration and the generated spectral bandwidth. The crystal is housed in an oven and maintained at a constant 100 °C. The OPO cavity is a standing-wave design comprising two plano-concave mirrors, $M_{1,2}$, and four plane mirrors, $M_3$-$M_6$. All mirrors are coated on 3-mm-thick $CaF_2$ substrate. The mirrors are all highly transmitting ($T$>90%) for the pump over 750–850 nm, highly reflecting ($R$>99%) for the signal over 1020-1600 nm, and have good transmission ($T$>80%) for the idler over 1850–5500 nm, thus ensuring SRO configuration for the OPO. All cavity mirrors are also chirped for dispersion control, with aGDD~0 over 1000-1200 nm, and varying from -500 to +200 $fs^2$ across 1000-1600 nm. A $CaF_2$ lens, $L_2$, is used to collect the mid-IR idler output, while a filter, F, separates the generated idler from the transmitted pump. The total round-trip optical length of the OPO cavity is ~1.5 m, corresponding to a repetition rate of 100 MHz, ensuring synchronization with the pump laser repetition-rate. Since the zero-GVD point in MgO:PPLNis at ~1900 nm, we use a pair of SF11 equilateral prisms, $P_{1,2}$, for intracavity dispersion control of signal pulses. While $M_5$ serves as the end mirror for the auxiliary cavity, $M_6$ serves as the end-mirror in the presence of dispersion compensation.

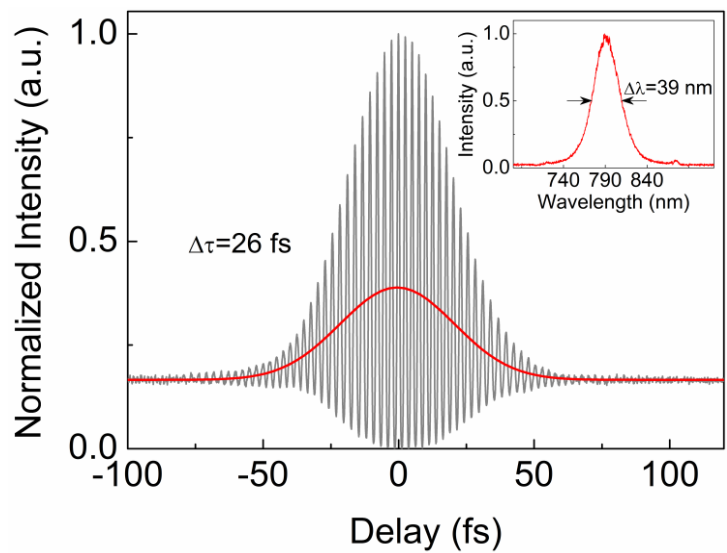

Fig. 2. Autocorrelation measurement of the Ti:sapphire pump laser. Inset: Corresponding spectrum centered at 790 nm.

## 3. Results and discussion

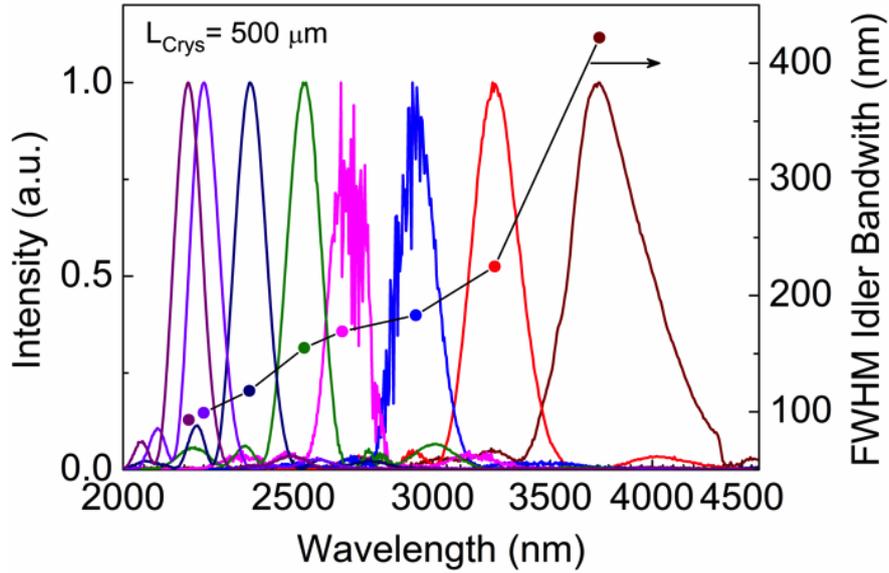

Fig. 3. Idler spectrum and variation of the FWHM idler bandwidth across the tuning range of the OPO.

In order to characterize the OPO, we initially performed spectral measurements of the mid-IR idler output across the tuning range. We chose an interaction length of 500 μm as a compromise between the nonlinear gain and the dispersion to achieve the broadest wavelength coverage. Wavelength tuning was achieved conveniently by cavity delay on either side of the perfectly synchronous cavity length. The spectra were measured using a commercial Fourier transform spectrometer, with the results shown in Fig. 3, where the abscissa is depicted in logarithmic scale. As evident from Fig. 3, the idler wavelength is continuously tunable from a center wavelength of 2179 to 3732 nm, over 1553 nm, with the spectra extending across 2000-4500 nm. The corresponding signal wavelength leaked out through one of the cavity mirrors is measured to tune from 1002 to 1239 nm, over ~24 nm. Also shown Fig. 3 is the variation of the idler spectral bandwidth across the tuning range. The FWHM idler bandwidth varies from 93 nm at 2179 nm to 422 nm at a center wavelength of 3732 nm. It is evident in Fig. 3 that the idler spectra exhibit stable, uniform, well-behaved, and smooth profiles across the full tuning range of the OPO, a characteristic of the SRO configuration. The spectral modulation in the ~2700-3000 nm region is caused by water vapor in the laboratory atmosphere.

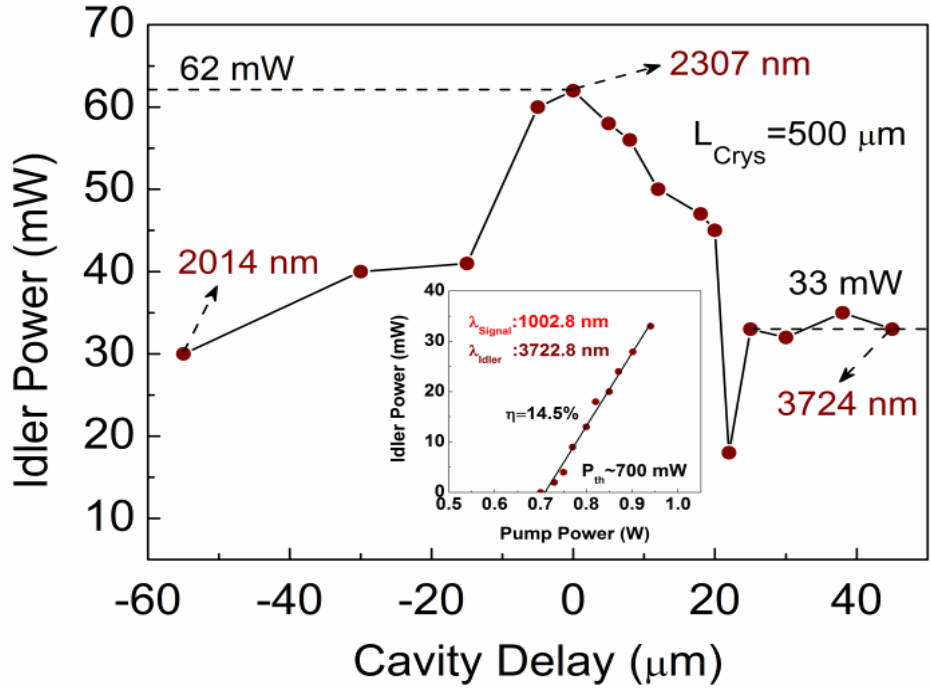

Fig. 4.Variation of the idler power as function of the cavity delay. Inset: Power scaling at an idler wavelength of 3723 nm.

We recorded the idler output power across the tuning range of the OPO for a fixed input pump power using the grating interaction length, $l$=500 μm. Figure 4 shows the variation of idler power as a function of cavity delay. Under perfect synchronization, we were able to extract up to 62 mW of idler power at central wavelength of 2307 nm.   As the cavity length is detuned from –55 μm, the idler power increases from 30 mW at 2014 nm, to a maximum of 62 mW at zero-detuning, beyond which it decreases to 33 mW at 3724 nm for 45 μm of detuning. Also shown in the inset of Fig. 4 is the variation of the idler power as a function of the pump power. At an idler wavelength of 3723 nm, corresponding to a signal wavelength of 1003 nm, we were able to generate 33 mW of idler power for a pump power of 940 mW at an extraction efficiency of 3.5% and a slope efficiency of 14.5%. At this significantly long wavelength, the threshold of the OPO is recorded to be ~700 mW.  The relatively high pump threshold could be attributed to the temporal walk-off between the pump and the signal pulses in the MgO:PPLN crystal. The group velocities of the pump and signal are calculated to be $v_{gp}$~$c/2.27$ and $v_{gs}$~$c/2.21$, respectively, where $c$ is the velocity of light. Given the input pump pulse duration of ~26 fs and the group velocities of the pump and signal pulses, the temporal walk-off length between the pump and signal is estimated to be $l_{eff}$~260 μm, beyond which the pump and signal pulses no longer overlap. This walk-off distance is about half the grating interaction  length of $l$=500 μm, implying that the grating does not fully contribute to parametric gain.

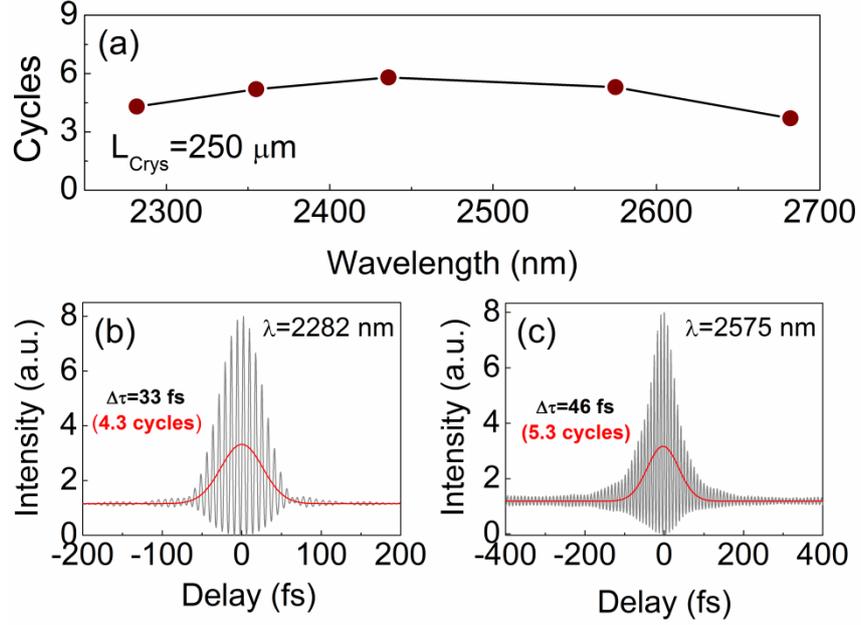

Fig. 5. (a) Optical cycles of the generated idler pulses using $l$=250 μm interaction length in MgO:PPLN. (b) and (c) Interferometric autocorrelation of the idler at 2282 nm and 2575 nm, respectively.

In order to exploit the largest spectral acceptance bandwidths associated with shortest interaction lengths, to generate the shortest pulse durations, we performed temporal characterization the output idler pulses using the grating interaction length of $l$=250 μm in the crystal. Under this condition, dispersion compensation was best optimized to generate idler pulses in the 2282-2682 nm wavelength range. Figure 5 shows the variation of the number of optical cycles as a function of the idler wavelength generated using the $l$=250 μm interaction length. As can be seen, we were able to generate few-cycle idler pulses across the measurement tuning range, ranging from 5.8 optical cycles at 2436 nm to as few as 3.7 optical cycles at 2682 nm. Also shown in the inset of Fig. 5 are two representative interferometric autocorrelation profiles of idler pulses at 2282 nm and 2575 nm, resulting in pulse durations of 33 fs and 46 fs, assuming $sech^2$ profile, corresponding to 4.3 and 5.3 optical cycles, respectively. At 2682 nm, where the dispersion compensation is well optimized, the measured FWHM idler bandwidth of 10.2 THz (245 nm) results in a time-bandwidth product of $\Delta\nu\Delta\tau\sim0.34$, close to the transform limit for an ideal $sech^2$ pulse, $\Delta\nu\Delta\tau\sim0.315$, confirming that the pulses are nearly chirp-free. Fine adjustment of dispersion compensation at other idler wavelengths could also result in similar near-transform-limited pulses across the tuning range. At some of the longer wavelengths, we observed distortion of the autocorrelation profile, which we believe to be due to water vapor absorption.

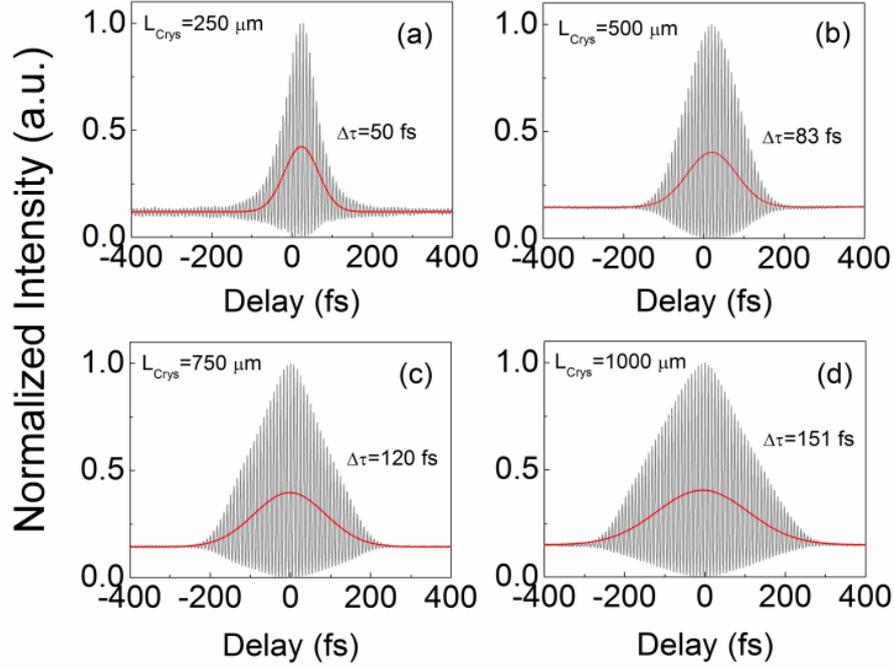

Fig. 6. Dependence of the idler pulse duration on the interaction length, (a) 250 μm (b) 500 μm (c) 750 μm and (d) 1000 μm of the MgO:PPLN crystal at a fixed operating wavelength of ~2400 nm.

We further investigated the effect of interaction length on the idler pulse duration at a fixed wavelength. Figure 6 (a-d) shows the measured interferometric autocorrelation traces of the idler pulses at ~2400 nm using $l$=250 μm, 500 μm, 750 μm, and 1000 μm grating interaction lengths in the crystal. As can be seen, the pulse duration increases with the increase in interaction length, associated with reduced spectral acceptance bandwidth, as expected. The variation in the idler pulse duration, and the corresponding spectral bandwidth, as a function of the interaction length is shown in Fig. 7. As the idler bandwidth (FWHM) decreases from 214 nm for $l$=250 μm to 60 nm for $l$=1000 μm, the pulse duration gradually increases from 50 fs to 151 fs at a fixed idler wavelength centered at ~2400 nm. Also shown in the inset of Fig. 7 is the corresponding time-bandwidth product, which is estimated to be constant, $\Delta\nu\Delta\tau$~0.48. This is close to $\Delta\nu\Delta\tau$=0.315 for an ideal $sech^2$ pulse shape, indicating that the generated idler pulses are nearly chirp-free and transform-limited for all grating interaction lengths. It should also be noted that the autocorrelation profile in Fig. 6(a) displays some distortion due to third-order dispersion, which we attribute to atmospheric absorption near ~2400 nm. Therefore, we expect even shorter pulse durations and lower time-bandwidth products $\Delta\nu\Delta\tau$<0.48 at other wavelengths across the idler tuning range.

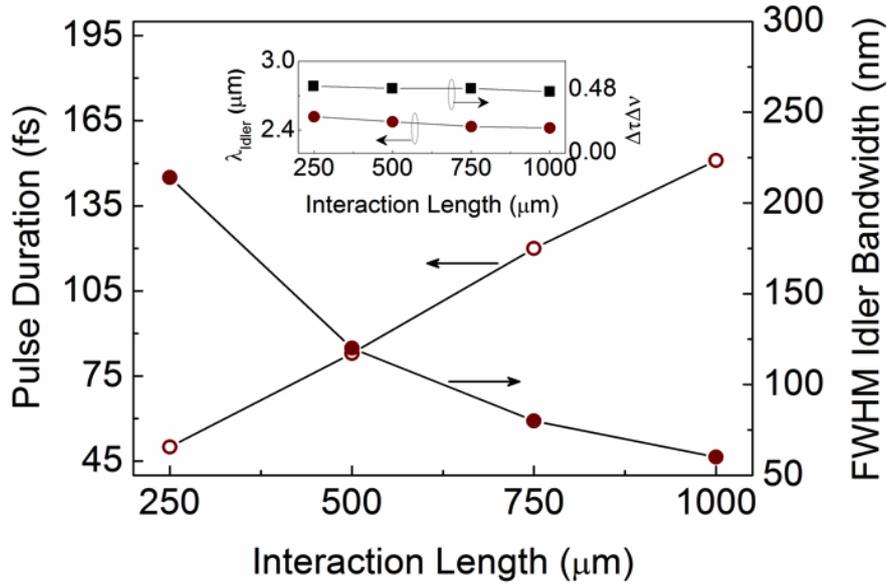

Fig. 7. Variation of the idler pulse duration and spectral bandwidth as a function of the interaction length at ~2400 nm. Inset: Idler wavelength and the time-bandwidth product of the idler pulses as function of the grating interaction length.

Finally, we investigated the long-term power stability of the idler output from the OPO at the longer mid-IR wave lengths and at maximum power. Figure 8 shows the result at an idler wavelength centered at 3652 nm. The measurements were performed with the Ti:sapphire pump laser repetition frequency locked to an RF reference by servo-control of the cavity length, but without active stabilization of the OPO cavity length itself. As can be seen, the idler is recorded to exhibit a power stability better than 1.6% rms over >2.75 hours, while operating at the maximum output power, with the standard deviation of the change in repetition frequency of the OPO as low as ~1 mHz centered at ~99.99 MHz. Also shown in the inset of Fig. 8 is the spatial quality of the idler output beam at 3762 nm, together with the orthogonal intensity profiles, confirming a single-peak Gaussian distribution in $TEM_{00}$ mode profile. We have repeated the same measurement at other wavelengths and found the idler to exhibit similarly high spatial beam quality across the tuning range.

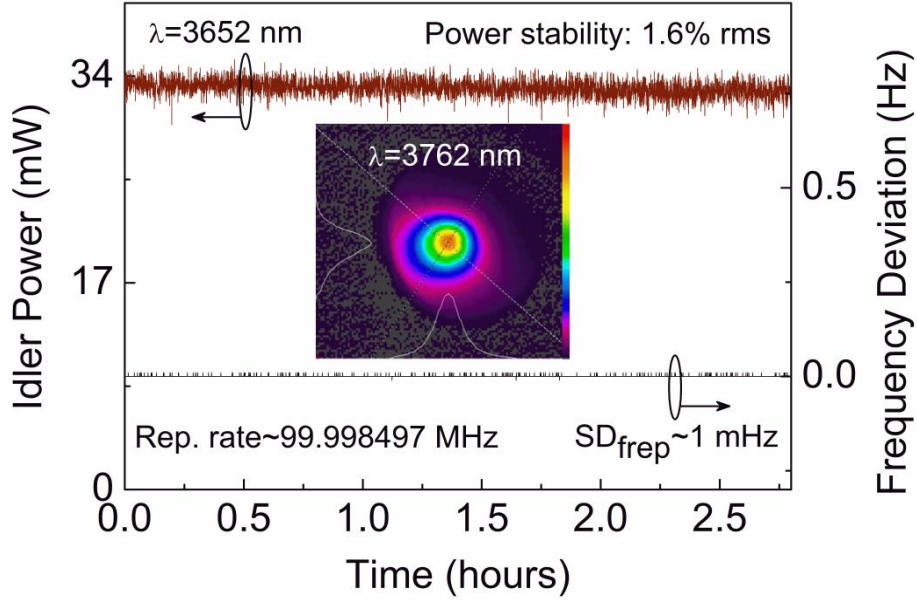

Fig. 8. Long-term power stability of the idler from the MgO:PPLN OPO and the standard deviation of the variation in the oscillator repetition frequency ($SD_{frep}$) over >2.75 hours. Inset: Spatial beam quality of the idler at 3762 nm.

## 4. Conclusions

In conclusion, we have demonstrated a singly-resonant femtosecond OPO for the mid-IR based on MgO:PPLN, synchronously pumped by a 20-fs KLM Ti:sapphire laser. The OPO is capable of delivering few-cycle broadband idler pulses with uniform and well-defined spectral structure and smooth tuning from a center wavelength of 2179 nm to 3732 nm, with a FWHM bandwidth of 422 nm at 3732 nm. The SRO configuration ensures continuous and uninterrupted tuning across the full wavelength coverage of the OPO by convenient cavity delay tuning. By using grating interaction lengths as short as $l$=250μm and careful dispersion control, we have generated near-transform-limited idler pulses of 3.7 to 5.8 optical cycles across 2282-2682 nm, with measurements at longer mid-IR wavelengths limited by the available diagnostics. We expect that the idler pulses at longer wavelengths will also be of a few optical cycles, and could be further controlled to achieve the shortest duration near the transform-limit by optimizing the intracavity dispersion compensation at shorter signal wavelengths. The OPO can provide >30 mW of output power over the entire mid-IR tuning range, with a maximum of 62 mW at 2307 nm. We have also investigated the effect of interaction length on the spectral bandwidth and pulse duration of the output from the OPO at a fixed idler wavelength, resulting in the generation of near-transform-limited pulses for all interaction lengths from $l$=250μm to $l$=1000 μm. With stabilization of the pump laser cavity length to a reference RF signal, thus minimizing the deviation of the repetition frequency down to ~1mHz, the OPO idler output exhibits a power stability better than 1.6% rms over >2.75 hours at 3652 nm, when operating at maximum output power, in excellent spatial beam quality with $TEM_{00}$ mode profile. Given the short interaction lengths of only 250 μm used in this OPO, with modified grating design and period, the spectral coverage may be further shifted into the mid-IR to wavelengths centered near 4500 nm, potentially extending to ~5000 nm, with reduced risk of idler absorption in the crystal. The described OPO represents a viable and practical source of few-cycle broadband radiation for applications in frequency metrology and spectroscopy in the mid-IR.

**Acknowledgements**

This research was supported by the Ministry of Science and Innovation, Spain, through project OPTEX (TEC2012-37853) and by the European Research Council (Project Multicomb, Advanced Investigator Grant 267854).

**References**

[1] T. W. Hänsch, "Nobel Lecture: Passion for precision," Rev. Mod. Phys., 78, 1297-1309 (2006).

[2] A. Schliesser, N, Picqué, and T. W. Hänsch, " Mid-infrared frequency combs," Nature Photonics **6**, 440-449 (2012).

[3] M. Ebrahim-Zadeh, "Mid-infrared ultrafast and continuous-wave optical parametric oscillators," in Solid-State Mid- Infrared Laser Sources, I. T. Sorokina and K. L. Vodopyanov (Eds.), Springer-Verlag Science Series, Topics Appl. Phys. **89**, 184-224 (2003).

[4] S. T. Wong, T. Plettner, K. L. Vodopyanov, K. Urbanek, M. Digonnet, and R. L. Byer, "Self-phase-locked degenerate femtosecond optical parametric oscillator," Opt. Lett. **33**, 1896-1898 (2008).

[5] F. Adler, K. C. Cossel, M. J. Thorpe, I. Hartl, M. E. Fermann, and Jun Ye, "Phase-stabilized, 1.5 W frequency comb at 2.8–4.8 μm," Opt. Lett. **34**, 1330-1332 (2009).

[6] K. L. Vodopyanov, E. Sorokin, I. T. Sorokina, and P. G. Schunemann, "Mid-IR frequency comb source spanning 4.4–5.4 μm based on subharmonicGaAs optical parametric oscillator," Opt. Lett. **36**, 2275-2277 (2011).

[7] A. Marandi, N. C. Leindecker, V. Pervak, R. L. Byer, and K. L. Vodopyanov, "Coherence properties of a broadband femtosecond mid-IR optical parametric oscillator operating at degeneracy," Opt. Express **20**, 7255-7262 (2012).

[8] N. Leindecker, A. Marandi, R. L. Byer, K. L. Vodopyanov, J. Jiang, I. Hartl, M. Fermann, and P. G. Schunemann, "Octave-spanning ultrafast OPO with 2.6-6.1 μm instantaneous bandwidth pumped by femtosecond Tm-fiber laser," Opt. Express **20**, 7046-7053 (2012).

[9] V. Ramaiah-Badarla, A. Esteban-Martin, and M. Ebrahim-Zadeh, "Self-phase-locked degenerate femtosecond optical parametric oscillator based on $BiB_3O_6$," Laser Photonics Rev. **7**, L55–L60 (2013).

[10] Z. Zhang, D. T. Reid, S. Chaitanya Kumar, M. Ebrahim-Zadeh, P. G. Schunemann, K. T. Zawilski, and C. R. Howle, "Femtosecond-laser pumped $CdSiP_2$ optical parametric oscillator producing 100 MHz pulses centered at 6.2 μm," Opt. Lett. **38**, 5110-5113 (2013).

[11] A. Stingl, M. Lenzner, Ch. Spielmann, and F. Krausz, "Sub-10-fs mirror-dispersion-controlled Ti:sapphire laser," Opt. Lett. **20**, 602-604 (1995).

[12] A. Esteban-Martin, O. Kokabee, and M. Ebrahim-Zadeh, "Efficient, high-repetition-rate, femtosecond optical parametric oscillator tunable in the red," Opt. Lett. **33**, 2650-2652 (2008).

[13] M. W. Haakestad, A. Marandi, N. Leindecker, K. L. Vodopyanov, "Five-cycle pulses near λ=3 μm produced in a subharmonic optical parametric oscillator via fine dispersion management," Laser Photonics Rev. **7**, L93–L97 (2013).